\documentclass[conference]{IEEEtran}

\usepackage{algorithmic, algorithm2e, color}
\usepackage{bbding}
\usepackage{amsmath, amssymb, mathtools}
\usepackage{float}
\usepackage{graphicx, epstopdf, url}
\usepackage{exscale,relsize}
\usepackage{amsfonts}
\usepackage{mdwlist}
\usepackage{xcolor}
\usepackage{paralist}
\usepackage[labelformat=simple]{subcaption}

\newcommand{\acro}{{\ensuremath{\mathcal{\small P{\tt o}P{\tt a}}}}\xspace}
\newcommand{\ignore}[1]{}

\title{{\em Assentication:} User Deauthentication and Lunchtime Attack Mitigation with Seated Posture Biometric}

\author{}
\author{
\IEEEauthorblockN{Tyler Kaczmarek}
\IEEEauthorblockA{University of California, Irvine\\tkaczmar@uci.edu}
\and
\IEEEauthorblockN{Ercan Ozturk}
\IEEEauthorblockA{University of California, Irvine\\ercano@uci.edu}
\and
\IEEEauthorblockN{Gene Tsudik}
\IEEEauthorblockA{University of California, Irvine\\gene.tsudik@uci.edu}
}
\begin{document}

\ignore{
\IEEEoverridecommandlockouts
\makeatletter\def\@IEEEpubidpullup{9\baselineskip}\makeatother
\IEEEpubid{\parbox{\columnwidth}{Permission to freely reproduce all or part
    of this paper for noncommercial purposes is granted provided that
    copies bear this notice and the full citation on the first
    page. Reproduction for commercial purposes is strictly prohibited
    without the prior written consent of the Internet Society, the
    first-named author (for reproduction of an entire paper only), and
    the author's employer if the paper was prepared within the scope
    of employment.  \\
    NDSS '18, 18-21 February 2018, San Diego, CA, USA\\
    Copyright 2018 Internet Society, ISBN 1-1891562-49-5\\
    http://dx.doi.org/10.14722/ndss.2018.23xxx
}
\hspace{\columnsep}\makebox[\columnwidth]{}}
}

\maketitle

\begin{abstract}
Biometric techniques are often used as an extra security factor in authenticating human users.
Numerous biometrics have been proposed and evaluated, each with its own set of benefits and pitfalls.
Static biometrics (such as fingerprints) are geared for discrete operation, 
to identify users, which typically involves some user burden. Meanwhile, behavioral biometrics (such as 
keystroke dynamics) are well-suited for continuous, and sometimes more unobtrusive, operation.
One important application domain for biometrics is {\em de-authentication}: a means of quickly
detecting absence of a previously-authenticated user and immediately terminating that user's active 
secure sessions. De-authentication is crucial for mitigating so-called {\em Lunchtime Attacks}, whereby
an insider adversary takes over (before any inactivity timeout kicks in) authenticated state of a careless user 
who walks away from her computer.

Motivated primarily by the need for an unobtrusive and continuous biometric to support effective 
de-authentication, we introduce \acro\ -- a new hybrid biometric based on a human 
user's seated posture pattern. % \acro\ offers unobtrusive continuous operation and a very short initial training period. 
\acro\ captures a unique combination of physiological and behavioral traits.
We describe a low-cost fully functioning prototype that involves an office chair instrumented
with $16$ tiny pressure sensors. We also explore (via user experiments) how \acro\ can be 
used in a typical workplace to provide continuous authentication (and de-authenication) of users. 
We experimentally assess viability of \acro\ in terms of uniqueness by collecting and evaluating 
posture patterns of a cohort of users. Results show that \acro\ exhibits very low false positive, 
and even lower false negative, rates. In particular, users can be identified with, on average, 
$91.0\%$ accuracy. Finally, we compare pros and cons of \acro\ with those of 
several prominent biometric-based de-authentication techniques.
\end{abstract}

\section{Introduction \& Motivation}
Secure, correct and efficient user authentication is an integral component of any meaningful security system. 
Authentication schemes implemented in a typical modern workplace typically include 
two factors: (1) a user demonstrates knowledge of a secret password or PIN, and (2) a user proves possession
of a secure device or token. However, it is becoming more popular to augment this approach with a third factor --
biometrics that reflect inherent human traits or behaviors. Biometric techniques are considered as 
the best means of evaluating human inherence and they range widely: from a very simple (e.g., fingerprints) 
to rather complex, such as iris scans.

After initial authentication, users often spend long stretches of time continuously using computing devices 
and services. During that time,  continuous presence of the originally authenticated user must be
periodically re-affirmed, especially, in a shared workplace setting. 
Failure to do so can result in so-called {\em Lunchtime Attacks}. Such an attack occurs when a previously 
authenticated user walks away from her workplace, thus allowing the adversary to take over her login 
session and engage in potentially nefarious activity. This prompts the need for periodic re-authentication and/or
continuous authentication. Unfortunately, the former can be quite annoying, as is the case with too-short inactivity 
time-outs requiring frequent password re-entry. 

Meanwhile, continuous authentication (or presence verification) 
is challenging in its own right. For example, camera-based methods that use face recognition 
\cite{chang2003comparison} or gaze tracking \cite{eberz2015preventing} might be viewed as intrusive in terms of 
personal privacy, since cameras can be abused (e.g., by malware) to surreptitiously 
record users. Furthermore, face recognition is prone to attacks, while gaze tracking requires the user to maintain 
line-of-sight with the camera, which can result in unnecessary de-authentication when the user turns away, 
while remaining at the workplace. Whereas, keyboard or mouse activity profiling and monitoring, though effective in 
some settings, are poorly suited for cases when a user temporarily halts input activity, e.g., in order to chat with 
co-workers or answer the phone. Other techniques continuously measure physical distance between the user and 
her workplace, by requiring each user to wear an extra device, e.g., a wristband or smart badge. Such methods
are: (1) potentially burdensome due to imposing an extra device, and (2) ultimately authenticate only the presence 
of the device and not of its owner.

Based on the above discussion, we believe that the ``design space'' for continuous authentication (or, 
equivalently, de-authentication) techniques needs to be explored further. From the outset, we acknowledge 
that a perfect continuous authentication method is unlikely to materialize; in fact, one might not even exist. 
In other words, since each previous method has a distinct set of advantages and limitations/flaws, the same 
will certainly hold for our current efforts. 

In this paper, we propose and evaluate a new biometric called \acro. It is based on a user's 
seated posture patterns in an average office chair over the course of a typical workday. 
We explore two potential use cases for this new biometric.
{\bf First}, we examine applicability of \acro\ to continuous user authentication, i.e., 
ensuring that -- after the initial successful login -- the person currently using a particular computer 
is the same as the one who initially logged in. One of \acro's key advantages over many other 
de-authentication methods is its ability to operate in an unobtrusive manner, with no effort on the 
part of the user.\footnote{This is in contrast with, for example, fingerprint-based continuous authentication, 
which would prompt the user to periodically swipe her finger(s) on the fingerprint reader; 
which is obtrusive and disrupts the typical workflow.} 
\ignore{
There have been other attempts to solve the problem of obtrusiveness through facial recognition, 
gaze tracking, or the human body's response to electrical pulses 
\cite{chang2003comparison,eberz2015preventing, rasmussen2014authentication} but 
these also come with their own set of flaws. For instance, facial recognition can be defeated by a 
photograph of a valid user, gaze tracking requires that a user's eyes are visible by the tracking system, 
and the pulse-response system requires that the user creates a complete circuit with their body and 
the evaluation system. All of these additional constraints serve to disrupt 
the typical user experience in the workplace, and are obtrusive to some degree. 
}

{\bf Second}, we evaluate \acro\ in the context of user identification. 
Unlike continuous authentication, user identification does not involve as much disruption of 
normal user behavior. However, it provides a different use-case for \acro. 
\ignore{
In the case of continuous authentication, it is appropriate for the \acro\ system to be aware of one user, 
perhaps the occupant of an office, or the owner of the chair itself. Authentication then becomes a question of 
"are you my owner or not?" Answering this question requires only a single user profile. Identification instead 
moves into a use case where many people may share this office or chair, and is more akin to a home setting, 
where the \acro\ system would need to store several separate user profiles and would instead need to answer 
the question "are you one of my owners, and if so: which one?"
} 
While continuous authentication requires periodic checks during a given session, identification is typically 
established at session initiation time and  requires faster operation. 

To evaluate its viability and effectiveness, we built a low-cost \acro\ prototype by instrumenting
a commodity office chair with ultra-thin flexible sensors that gather user posture data. 
Its purpose was to assess whether users are correctly authenticated, based on their own training data. 
The same platform was used to test: (1) uniqueness of \acro\ within a sample population of measured 
users, and (2) its stability over time, i.e., permanence. Our results demonstrate that
the prototype unobtrusively captures the necessary data for continuous authentication and identification
while the user engages in a typical use of a desktop or laptop computer.

The rest of this paper is organized as follows: Section \ref{sec:rw} overviews related work. Next, 
Section \ref{sec:bg} provides the background on continuous authentication and de-authentication. 
Section \ref{sec:bio} describes the \acro\ biometric. Then, Section \ref{sec:threat} outlines the adversarial 
model, followed by Section \ref{sec:comparison} which compares several prominent biometric-based 
continuous authentication methods. Section \ref{sec:method} describes the \acro\ prototype and 
methodology used for data collection, followed by results in Section \ref{sec:results}  and a discussion of 
their implications in Section \ref{sec:discuss}. The paper concludes with future work directions in Section \ref{sec:FW} and a 
summary in Section \ref{sec:conc}.

\section{Related Work}
\label{sec:rw}
Biometric traits have been extensively explored in the context of authentication. 
Jain et al. \cite{jain2006biometrics} provides an authoritative overview of many well-known 
techniques, including: fingerprint, face, iris, palm-print and keystroke dynamics. However, since 
our focus is on biometric-based continuous authentication which can be used to achieve 
effective de-authentication, we do not discuss methods that are not amenable for the intended application.

Rasmussen et al. \cite{rasmussen2014authentication} use human body's response to electric signals 
as a biometric. In the proposed system, a weak pulse signal is applied to the palm of one hand and 
measured on the palm of the other hand. Pulse-response biometric can be used as a second or third 
factor in user authentication and/or as a continuous authentication mechanism. The system achieves 
100\% accuracy over a static set, and 88\% accuracy on permanence tests performed over several weeks.

Eberz et al. \cite{eberz2015preventing} investigate eye movement patterns as a biometric. Based on 
gazing data gathered from 30 participants, pupil, temporal and spatial features are defined. Reported 
equal error rate is 3.98\% in a single session and 92.2\% of attacks are detected within 40 seconds. 
Measurements done two weeks apart show that this biometric is stable over time. 

Mare et al. \cite{mare2014zebra} propose wearing a bracelet that has a gyroscope and an accelerometer 
for continuous authentication. When the user interacts with the computer (e.g., typing or scrolling), the 
bracelet transfers collected sensor data to the computer, which evaluates whether user actions match 
sensor data. The proposed system, ZEBRA, achieves continuous authentication with 85\% accuracy 
and detects attacks within $11$ seconds. However, a recent study by Huhta et al. \cite{huhta2016pitfalls} 
presents a set of credible attacks on ZEBRA.

ECGs and EEGs have been used to construct continuous authentication systems. \cite{louis2016continuous} 
uses ECG and a novel feature extraction method called One-Dimensional Multi-Resolution Local Binary 
Patterns (1DMRLBP). Results show a false positive rate of 0.39\% and a false negative rate of 1.57\%. 
The system uses ECG data gathered  mostly for medical purposes and it is unclear how this data can 
be obtained in a realistic environment. Also, Riera et al. \cite{riera2008starfast} use both ECG and EEG 
signals for continuous authentication, achieving a false negative rate of 0.82\%  using a multistage fusion 
architecture to combine authentication results from ECG and EEG signals. In order to gather 
EEG signals, this system requires each user to wear a special cap that covers the head.

Keystroke dynamics are another means of continuous authentication. \cite{ahmed2014biometric} uses 
$1,500$ digraphs from each user as a base profile and applies neural networks to guess missing 
digraphs. In a 53-user experiment, a false negative rate of 0.0152\%, a false positive rate of 4.82\% 
and an equal error rate of 2.46\% are achieved.

Finally, Conti et al. \cite{conti2016fadewich} describe FADEWICH, a continuous authentication system 
that uses attenuation of wireless signals when a human body is on the signal's path. FADEWICH is 
{\bf not} based on any biometrics. It tracks the user by placing 9 sensors in a 6m-by-3m office environment.  
Once detected as having left the environment, the user is logged out. FADEWICH successfully 
de-authenticates users with 90\% accuracy within 4 seconds, and 100\% accuracy within 6 seconds.

There have been prior attempts to use posture and seated pressure for both identification and 
continuous authentication.  Gia et al. \cite{gia2012people} used data gathered from: (1) four pressure 
sensors placed on the seat bottom, (2) an accelerometer, and (3) light sensors placed on the seat-back, 
to identify the user. Pressure sensors are used to differentiate among users, while weight and 
accelerometer readings determine chair movements when someone sits down. Also, light sensors 
determining how much space is covered by the sitting user. In an experiment involving only 10 people, 
a rather low accuracy of 72\% is achieved.

Furthermore, Yamada et al. \cite{yamada2009soft} described a hip-print authentication method 
which uses pressure data from 32 sensors placed along the seat bottom. Since any seated shift 
action causes a short continuous change of pressure values, two authentication methods are considered: 
one based on stable state of pressure distributions, and another based on the transient part. In an 
experiment with 12 participants, accuracy of 96.3\% is achieved for the former, and 92.9\% 
-- for the latter. However, in experiments that use only first three pressure measurements (1.5 seconds), 
quite low accuracies of 74.3\% and 59.9\% are reported for $10$ and $25$ subjects, respectively.

Among prior work, the one closest to this paper is \cite{riener2008supporting}. It proposes a continuous 
driver identification system for automobiles that uses pressure data from two mats, each containing $32$x$32$
sensors, placed on the seat cushion and backrest of the driver seat. Features used for classification are 
based on one's pelvic bone signature, mid- to high-pressure distribution and weight.  In a study involving 
34 participants, a fairly low uniqueness rate is reported. The experiment was done using only one 
posture, which makes it hard to identify posture characteristics for different people. Moreover, a car 
setting is not appropriate for detecting pressure distribution changes, since most drivers adopt one 
constant posture adjusted to their preferred driving position.

Finally, Mutlu et al. \cite{mutlu2007robust} investigated how to use fewer sensors to detect posture. 
To determine optimal sensor placement, a classifier is constructed that learns the probabilistic 
model between the chosen subset of sensor values and feature vectors used for posture classification. 
With $19$ sensors, classification accuracy of 87\% is reported.  As discussed later, this study guides 
our sensor placement strategy.

\section{Background}
\label{sec:bg}
This section sets the stage for the rest of the paper by overviewing user authentication, 
de-authentication, attack scenarios and continuous authentication requirements.

\subsection{User Authentication}
User authentication can involve one or more of the following factors: 
\begin{compactenum}
\item[\framebox{F1:}] What one knows, or what one recognizes.\\
The former corresponds to knowledge of: passwords, PINs, drawing patterns and free-text 
answers to security questions. The latter corresponds to recognition of: correct answers to 
multiple-choice questions, faces or other types of images.
\item[\framebox{F2:}] What one has in their possession.\\
This generally means some form of a personal (even passive) device, such as a 
badge, bracelet, key-fob, token or smartphone. 
\item[\framebox{F3:}] What one is, or how one behaves.\\
The former type is referred to as a {\em static} and includes biometrics based on: 
fingerprints, irises, palms, wrists, faces, ears and pulse-response.
The latter type is called {\em behavioral} and includes biometrics based on:
gait, keystroke dynamics, head movements, hand gestures and gaze tracking.
\end{compactenum}
Though widely used, F1-type authentication alone is widely considered to be insufficient, mainly due 
to the low entropy of secrets involved. By itself, F2 is also inadequate, since a personal device is not 
guaranteed to always be in possession of its intended owner. Finally, F3 can be subverted, at least for 
some static methods, e.g., via cloned fingerprint moulds \cite{uludag2004attacks}, fake irises using 
contact lenses \cite{bowyer2014cosmetic}, and face masks \cite{irisSpoof}. It also often requires a 
non-trivial training or enrollment phase. Meanwhile, some behavioral biometrics are unstable or fragile, 
e.g., gait and head movements.  Consequently, multi-factor user authentication is usually recommended 
in order to achieve better security.

\subsection{De-Authentication \& Lunchtime Attacks}
As part of everyday office or workplace activity, an average user might engage in one or more 
of the following activities (not an exhaustive list): 
\begin{compactenum}
\item[\mbox{[A1]:}] Work by continuously utilizing one or more traditional input devices, 
	such as a keyboard, touchscreen or mouse.
\item[\mbox{[A2]:}] Take a quick seated nap or meditation break.
\item[\mbox{[A3]:}] Read some printed matter, e.g., a paper or book.
\item[\mbox{[A4]:}] Use another personal device, e.g., a smartphone.
\item[\mbox{[A5]:}] Turn away from one's desk to talk to other people directly, or on the phone.
\item[\mbox{[A6]:}] Watch videos and/or listen to music without using any input devices.
\item[\mbox{[A7]:}] Take part in an audio or video conference.
\item[\mbox{[A8]:}] Get up momentarily to fetch something from the immediate vicinity 
	(or simply to stretch) and return.
\item[\mbox{[A9]:}] Walk away from the workplace for a short (e.g, bathroom), longer (e.g., lunch),  or 
	long (e.g., done for the day) time, before returning.
\end{compactenum}
In a security-conscious setting, these activities might require periodic reassurance that the same user 
(who initially authenticated and/or logged in) is still present. Ideally, when the original user
remains present [A1-A7], no reassurance should be needed. However, [A9] results in leaving the 
workplace unattended, while [A8] might. (Also, [A2] could be viewed as the user not really being there.)
An important challenge is to distinguish among these types 
of activities. The term {\em de-authentication} denotes the process of deciding whether the original
user is no longer present and, if so, terminating active secure sessions.

In a perfect world, each user would always log out or otherwise terminate all active sessions before 
stepping away. Unfortunately, this is far from reality, which triggers the threat of {\em Lunchtime 
Attacks}. As the name suggests, attack of this type occurs when the adversary
takes over the secure session(s) of a legitimate user who has left, even for a short time.
Such attacks are quite common, as noted in the recent work of Marques et al. \cite{snooping}.

\subsection{Default Approach: Inactivity Timeouts}
The most common current means of dealing with Lunchtime Attacks and reassuring original
user presence is inactivity timeouts. Most users of personal and workplace
computing devices are familiar with them: whenever keyboard and/or mouse inactivity 
exceeds a certain threshold, de-authentication takes place, i.e., log-in and other (previously
authenticated) sessions are terminated. Various operating systems, apps and websites set 
their own timeout rules and policies. In some cases (e.g., macOS or Windows) users can 
select their own timeouts. At a typical workplace, mandatory timeouts are often imposed.

Saying that timeouts are unloved would be an understatement. As noted in \cite{mare2014zebra},
most users find too-short timeouts annoying, while too-long timeouts are insecure, since they
defeat the purpose of Lunchtime Attack mitigation (by extending the attack time window).
Even more importantly, timeouts achieve their desired effect only in case [A1]
and fail in several other ways:
\begin{compactitem}
\item They operate under the assumption that keyboard/mouse inactivity (i.e., "NOT [A1]") 
indicates user absence. This is often not true, e.g., in cases [A2]-[A5] and [A8]. De-authenticating 
the user in these cases is both unnecessary and annoying.
\item Conversely, timeouts na\"ively suppose that resumption of activity (within the timeout threshold)
indicates presence of {\em the same} user. This is clearly wrong in situations where the original walks
away [A9] and the adversary quickly starts typing. 
\item In case [A6], if timeouts are activated, the user is also unnecessarily burdened. Otherwise, if 
timeouts are automatically disabled while music and/or videos are playing, the user can walk away for 
a potentially long time, thus leaving the computing device(s) open to Lunchtime Attacks.
\item The same holds for case [A7], except that user's voice and/or camera movements might be used to 
infer continuous presence. However, this would require additional voice or visual authentication.
\item In case [A9], timeouts only work correctly (by de-authenticating the original user) if no
attack occurs. Knowing the timeout threshold, which is usually not secret, allows the adversary to 
easily succeed in a Lunchtime Attack.
\end{compactitem}

\subsection{Continuous Authentication}
Given the inadequacy of inactivity timeouts, one appealing alternative is continuous authentication.
Methods of this variety are generally unobtrusive, i.e., require none or very little user burden.
As discussed in Section \ref{sec:rw}, these include: keystroke dynamics \cite{ahmed2014biometric}, 
wrist movement \cite{mare2014zebra}, pulse response \cite{rasmussen2014authentication}, 
gaze tracking \cite{eberz2015preventing}, and wireless signal monitoring \cite{conti2016fadewich}.
(Note that only the first four are biometric-based methods, while the last is purely a de-authentication
technique.)

Keystroke (and/or mouse) dynamics are an effective and unobtrusive means of continuous authentication.
An adversary who steps in once the legitimate user leaves is likely to be quickly detected.
However, this method has the same pitfalls as inactivity timeouts: [A2]-[A8] can be either
interpreted as user's absence (which could be wrong) or as presence, which could also be wrong
if what actually occurred is [A9]. 

Wrist movement monitoring (e.g., using an accelerometer and a gyroscope contained in a 
bracelet as in \cite{mare2014zebra}) has much the same flaws as keystroke dynamics. Moreover, recent 
results \cite{huhta2016pitfalls} demonstrated a worrisome weakness of this approach.

Pulse response biometric \cite{eberz2017evaluating,rasmussen2014authentication} measures human body 
impedance, when subjected to a weak electrical signal. Obtaining measurements (hand-to-hand) requires 
closing a circuit which can be attained when the user types with both hands on conductive keyboard, 
uses both a keyboard and a mouse, or otherwise touches a conductive surface with each hand.
Once again, the limitations are similar to those of keystroke dynamics. (The only exception is that
pulse response works even if the user simply rests her hands on conductive surfaces while not typing
or moving the mouse). 

Gaze tracking \cite{eberz2015preventing} 
is also partially effective. It operates correctly, particularly in cases [A1], [A6] and [A7], 
as long as the user looks in the general direction of the camera. The same holds for [A9], and 
possibly [A8], as long as the gaze tracking interruption threshold 
is set appropriately. However, in cases [A2] and [A5], gaze tracking would incorrectly de-authenticate the user.
Directional (line-of-sight) nature of gaze tracking complicates [A2], while [A5] is simply not addressable.
Whether gaze tracking works in [A3] and [A4] also depends on whether line-of-sight is preserved.

Finally, the recent FADEWICH de-authentication technique \cite{conti2016fadewich} requires instrumenting 
the single-user workspace with multiple wireless devices that continuously measure WiFi
RSSI\footnote{RSSI: Received Signal Strength Indicator.} in order to detect whenever the user
exits her work-space. Since no external device is required, this places no personal burden on the user. 
On the other hand, by monitoring RSSI fluctuations, it is not possible to identify or 
authenticate the user. Also, this method can produce incorrect results (both false positives and 
false negatives) when multiple users are present and one leaves.

\subsection{Design Goals \label{goals}}
Since our main goal is the design of a biometric-based de-authentication method, we first 
consider general design goals for biometrics. A popular survey of biometric techniques by 
Jain, et. al \cite{jain2006biometrics} provides a comprehensive overview of many popular 
methods, and discusses design criteria, which include the following:

\emph{Universality:} The biometric must be (ideally) universally applicable. For example, 
an iris scanner is not useful for users who are missing an eye or have cataracts, while 
fingerprint readers are similarly useless for people with severe eczema.

\emph{Uniqueness:} The biometric must be unique within the target population. 
It must be possible to distinguish users using the biometric.

\emph{Permanence:} The biometric must remain consistent over the intended use duration. 
Not all biometrics are truly constant over a person's entire adult lifetime. 

\emph{Unobtrusiveness:} The biometric should be maximally transparent. Ideally, it should
be used in a passive manner, without any extra requirements or interference with users' normal behavior.

\emph{Circumvention Difficulty:} To be meaningful in any security context, the biometric must be difficult to circumvent. 
That is, false positive (fraud) rate should be minimal, i.e., it should be hard to impersonate a genuine user.

\emph{Low Error Rate:} The biometric must have a low false negative (insult) rate, i.e., should very rarely fail to 
recognize an enrolled user.

\emph{Collectability:} The biometric should be measurable in an fast, easy and meaningful quantitative way. 

\emph{Cost Effectiveness:} The biometric's distinguishing power as related to the cost of 
deployment and maintenance. In our design, this is a key goal. 

\emph{Easy enrollment:} The biometric's initial (training) phase should be as short and burden-free as possible.

\emph{Acceptability:} The ideal biometric is one which (most) users are comfortable to use.

We now present design goals for an ideal de-authentication method, not necessarily based on 
biometrics.\footnote{We do this while keeping in mind that all of them are unlikely to be achievable.}
\begin{compactitem}
\item Minimal extra components (particularly, physical or hardware) and monetary cost
\item Quick and correct detection of activities requiring de-authentication, i.e., [A9] or a 
circumvention attempt, e.g., another user sits down
\item Minimal false negative rate, i.e., probability of mistaking [A1]-[A8] for [A9]
\item Minimal false positive rate, i.e., probability of mistaking [A9] for [A1]-[A8] 
\item Maximal user transparency, i.e., unobtrusiveness 
\end{compactitem}
We recognize that the last goal might be ethically dubious. De-authentication methods with 
user transparency can be abused, e.g., by unscrupulous employers, to surreptitiously spy 
on unsuspecting users. We acknowledge that it is very 
difficult to reconcile positive and negative connotations.

\section{\acro Biometric}
\label{sec:bio}
\acro posture pattern biometric works by monitoring, over time, changing pressure patterns exerted by a 
user seated in a typical office chair. \acro relies on a combination of user behavioral 
patterns and physical characteristics. The latter  includes: hip width, spine length, leg length, torso width, as 
well as overall weight. In addition, overall pressure distribution and its shifts are determined by the user's exact 
posture. Over time, it changes in a way that is unique to each user and that user's emotive state. This behavioral 
characteristic is also factored into \acro.

\subsection{Strengths \& Weaknesses}
Since exact distribution of seated pressure depends on the user's physical dimensions as well as on 
adopted postures, \acro\ is a hybrid biometric blending physiological and behavioral factors. 
This allows it to benefit from some strengths of both. In particular, one's posture pattern can be 
captured in a strictly passive manner. Even though this property is 
shared by other biometrics, such as facial recognition or pulse response, posture pattern is not easily 
circumventable (unlike, e.g., facial recognition), and does not alter normal user behavior, unlike, e.g., 
pulse-response. We believe that this combination of unobtrusiveness, difficulty of circumvention, and behavior 
agnosticism make \acro\ an attractive biometric. 

However, posture pattern also inherits some weakness of both types of biometrics. As in any 
physiological biometric, it it impossible to capture one's template (i.e., posture pattern) without the 
use of hardware specifically instrumented for this task. Fortunately, \acro\ requires very little in terms of 
specialized hardware. As discussed later in Section \ref{sec:method}, we constructed an \acro\ prototype 
of an instrumented office chair.  Also, similar to many behavioral biometrics, permanence of \acro\ is not 
ideal. For example, a user who has a leg, hip or lower-back injury might appreciably alter her posture pattern. 
%Implications of such changes are discussed in Section \ref{sec:discuss}. 
%%GTS: last sentence: WE DO NOT DISCUSS this in that section!!! Why? Did you mean Section \results?
%%TK: I am not sure what that last sentence is saying, but it does not contribute to anythign, so im redacting it
\subsection{Liveness \& Replay}
In any biometric system used for continuous authentication, liveness detection is a serious concern. For example, 
a face recognition system needs to detect blinking, breathing, and/or some other artifact of a user being alive and 
present. Otherwise, as has been demonstrated in the past, it can be subverted by a photo or a mask (face-cast).
Traditionally, liveness is attained via some form of a challenge by the system that requires the user to act.
In case of facial recognition, the system might prompt the user to turn her head or look in a particular direction. 
While this helps achieve liveness and protect against subversion, it also sacrifices transparency and increases user
burden. 

Some modern de-authentication systems, such as gaze tracking or keystroke patterns, can passively 
check for liveness 
by relying on dynamic user behavior instead of constant physical characteristics. However, they require the user to 
act in a particular (not necessarily free or natural) manner. For example, gaze tracking requires the user to face 
in the general direction of the gaze tracking apparatus, which may not always be in the user's typical workflow. 
Furthermore, gaze tracking requires the user's eyes to be open. In the same vein, keystroke analysis requires the 
user to type on the keyboard. For its part, the pulse-response biometric needs the user to complete an electrical 
circuit by touching conductive implements with both hands. With all these systems, if the user fails to behave in 
the required manner, the likely outcome is a false positive.

In contrast, \acro\ is more forgiving in such cases. Furthermore, \acro\ does not rely on specific user actions, 
unlike implicit behavioral requirements of gaze tracking, keystroke analysis or pulse response. Instead, \acro 
is based merely on user's physical presence. It monitors seated pressure distribution regardless of 
whether the user faces the workstation, touches the keyboard with both hands, is currently typing, or keeps
eyes open. The only requirement for collection of posture pattern data is that the user must be 
seated in the chair. We believe that this makes \acro\ a good candidate for both 
continuous authentication and de-authentication.

\section{Adversarial Model \& Attacks}
\label{sec:threat}
\acro\ biometric focuses on protecting against insider threats. We are particularly
concerned with aforementioned Lunchtime Attacks whereby the adversary steps in to 
access a co-worker's computer after the latter walks away. 
Insider threats are not limited to such attacks, and might include scenarios ranging from a disgruntled employee 
staying after hours to sabotage a colleague, to the trivial case of a user deliberately giving access to a co-worker. 
In all scenarios, the adversary ``wins'' by gaining access to secure log-in or application sessions. 

We assume that the original user provides authentic log-in credentials at session initiation time.
However, the same user neglects to log-out before physically leaving the workplace. Once
the original user leaves, the adversary approaches the computer, accesses secure log-in
sessions and performs some actions, e.g., copy or erase sensitive files, read or send private email.
Such attacks are particularly dangerous since they originate from valid and logged-in 
user accounts. Also, it might be very difficult for the victim to repudiate the adversary's actions. 

Insider attacks are unfortunately quite commonplace. in fact, they account for about $28\%$ of all
electronic crimes in industry \cite{mickelberg2014us}. This includes some high-publicity attacks, 
such as the infamous 2014 Sony hack \cite{robb2014sony}.

We consider two types of insider adversaries:  \emph{casual} and \emph{determined}. 
In both cases, the adversary is aware of \acro's use and presence. The adversary is considered successful if it
manages to circumvent the system, either by physically imitating the victim's pressure patterns, or by constructing
an accurate model (replica) that does the same. We assume that the adversary cannot disable the 
system, or interfere with its correct operation through physical sabotage, since such manipulation would leave 
traces. 

The \emph{casual} adversary aims to subvert \acro\ through behavioral imitation of the victim user's posture patterns. 
We assume that this adversary is familiar with the habits and schedule of the victim, and has physical access to the 
victim's workplace.  Success of the casual adversary relies on the discriminating power of the system. In our prototype
design (discussed later), posture pattern data is aggregated and evaluated against the previously constructed profile 
every 10 seconds. Even in the unrealistically ideal scenario where the casual adversary instantly appears in the victim's 
chair immediately after the victim walks away, only 10 seconds would remain to perform any attack. (In our experimental office
setting, this attack time window is substantially shorter, $\approx$2-to-4 seconds, since it takes 3-to-4 seconds for the 
victim to leave and about as long for the adversary to enter and sit down.) After that, posture 
data would be flagged as incorrect, the victim would be de-authenticated, i.e., all active secure sessions would be terminated.

The \emph{determined} adversary seeks to defeat the system by fabricating a physical model of the victim user. 
We assume that this adversary has access to the exact sensor data of the victim, as well as precise measurements of 
the victim's posterior and lower back. This data might be obtained if the adversary manages to previously trick the victim 
to sit (for a sufficiently long period) in a staged chair instrumented the same way as the victim's. 

A perfect mold or cast of the victim with the correct pressure distribution would circumvent \acro. However, creation 
and deployment of such a mold is not trivial. The determined adversary would  have to create (perhaps using a 3D printer) 
a bulky and heavy object that accurately replicates the victim's posterior as well as lower back and weighs enough to exert the 
necessary pressure upon the instrumented chair, in the right places. Physically and logistically, deploying the mold
onto the victim's chair is burdensome and likely to be detected by extraneous means, e.g., the adversary might be
seen by others, or recorded by a security camera, while lugging the conspicuous mold.

However, we recognize that a mold is not the only way to subvert \acro. We conjecture that a more 
effective and discrete approach is to use a set of strategically placed hydraulic or pneumatic contraptions, 
each calibrated to exert an accurate amount of pressure on each sensor on the victim's chair. 
This kind of precision is difficult to achieve and, unlike a monolithic mold, placing
the entire set of contraptions onto the chair at the same time is also quite hard.
All in all, we consider this attack to be quite improbable and close to the realm of ``Mission: Impossible''.

\section{Comparison of Continuous Authentication Techniques}
\label{sec:comparison}
This section provides a comparison of five modern implementations of continuous authentication techniques,  
in the context of our design goals presented in Section \ref{goals}. We assume that systems would be 
deployed in a medium-sized office setting with about 50 employees. We included:
gaze tracking, pulse-response, wrist movement (ZEBRA), keystroke dynamics and \acro.
Each system is evaluated separately below and a summary is reflected in Table \ref{tab:comp}.

\subsection{Gaze Tracking}
We use the method proposed by Eberz, et. al \cite{eberz2015preventing} which describes a system for 
continuous authentication based on the Eye Movements biometric, which tracks user's gaze throughout 
a typical workday.

\emph{Universality} is (mostly) satisfied since it is reasonable to assume that a typical computer user has both eyes unobstructed
by dark glasses or severe cataracts. However, as mentioned earlier, activities [A2], [A5] and [A8] disrupt gaze tracking 
and are likely to result in unnecessary de-authentication. [A4] [A6] and [A7] might result produce the same outcome.

\emph{Uniqueness} was demonstrated experimentally by the construction of a classifier that distinguished among 30 
enrolled subjects with equal error rate of $3.98\%$.

\emph{Permanence} is a challenge for gaze tracking. User data relies on many environmental variables, such as 
overall brightness of the area, as well as on user-specific variables, such as user's overall level of fatigue. 
Such confounding variables make gaze tracking a somewhat ephemeral measure. 

\emph{Unobtrusiveness} is mostly achieved since gaze tracking {\em passively} collects eye movement data 
and does not require any special user input from the user {\bf as long as} she is facing the gaze tracking device.

\emph{Circumvention Difficulty} is framed in the context of an insider adversary attempting a lunchtime attack. 
A given user's gaze tracking pattern cannot be imitated by another with any degree of reliability.

\emph{Collectability} is achieved by the definition of 21 unique features of the human eye gazing at a computer 
screen. These features are discretized and quantitatively evaluated. 

\emph{Cost Effectiveness} represents another challenge. While this biometric appears to have high distinguishing power, 
based on difficulty to circumvent, permanence and uniqueness, high-fidelity, high-frame-rate cameras used in the
prototype implementation are quite expensive for large-scale deployment. The original prototype used the SenseMotoric 
Instruments (SMI) RED 500Hz gaze-tracking camera. While RED500 is no longer available, SMI has a range of 
gaze-tracking techniques that range from \$2,000 to \$5,000 per unit\footnote{see: \url{https://www.smivision.com}}. 
Even with a generous reduction of price for ordering high fidelity cameras in bulk, we estimate that the cost for instrumenting $50$ workstations to use a gaze tracking system would be on the order of $\$100,000.$  
\subsection{Pulse-Response}
%!TEX encoding = UTF-8 Unicode
Rasmussen, et. al \cite{rasmussen2014authentication} proposed a  biometric technique for continuous user authentication 
based on the human body's unique response to an electrical pulse. A user holds two diodes (one in each hand) a 
weak electrical current (1V) is passed through and measured.

\emph{Universality} is satisfied since the only requirement for users is to have two hands. 

\emph{Uniqueness} is provided using a classifier that matches pulse-response data to samples 
taken immediately beforehand, with nearly $100\%$ accuracy.

\emph{Permanence} is experimentally asserted to be very strong. Impedance was measured several times
over several days, and measurements were classified with the accuracy nearing $100\%$.

\emph{Unobtrusiveness} is fair, though not as good as that of gaze tracking, since for measurement purposes, 
the user needs  to touch two conductive surfaces with both hands. This imposes a small burden on the user and 
requires slight behavior modification. (Unless the user is constantly typing with both hands touching a conductive 
keyboard, or one hand on keyboard and the other -- on the mouse.) 
\emph{Circumvention Difficulty} is achieved since it is highly improbable for the adversary to mimic someone 
else's impedance, due to experimentally confirmed strong uniqueness and permanence features.

\emph{Collectability} is easily obtained, albeit additional instrumentation is needed, e.g., conductive keyboards, mice
and potentially other implements. 

\emph{Cost Effectiveness} of pulse-response is unknown. Although Rasmussen et al. mention possible 
realizations, no realistic prototype exists. Thus, it is impossible to estimate costs.

\subsection{Wrist Movements (ZEBRA)}
ZEBRA is a continuous authentication technique that relies on a wristband equipped with an accelerometer, 
gyroscope and radio to measure user's dominant wrist movements \cite{mare2014zebra}.

\emph{Universality} is satisfied since data collection requires only that the users {\em have} a dominant wrist, and
wear the wristband on it.

\emph{Uniqueness} is provided since it is claimed that  ZEBRA can identify: (1) a genuine user with 85\% accuracy, within 11 seconds. 
and (2) an adversary, within 50 seconds. 

\emph{Circumvention Difficulty} is dubious in light of recent attacks that violate ZEBRA's \emph{Uniqueness} features. 

\emph{Permanence}  is demonstrated by the classifier's ability to identify the correct user (within 11 seconds with $85\%$ accuracy). 
This holds for user sessions occurring several days apart.

\emph{Unobtrusiveness} is double-sided: on one hand, user behavior is mostly unmodified for the purpose of authentication; 
on the other hand, the user is burdened with wearing an additional device.

\emph{Collectability} is satisfied since wrist movement data is easily collected via gyroscope and accelerometer and 
communicated over Bluetooth.

\emph{Cost Effectiveness} is relatively good. Inexpensive fitness-style bracelets (equipped with an accelerometer, gyroscope, 
and Bluetooth) are available on the market at about $\$100$ per unit\footnote{see: \url{https://www.fitbit.com}}, making 
for a total of $\$5,000$ to instrument a medium-size ($50-$-person) office.

\subsection{Keystroke Dynamics}
Many continuous authentication systems based on keystroke dynamics have been proposed. We refer to 
\cite{ahmed2014biometric} as a representative example.

\emph{Universality} is satisfied since the use of a keyboard is routine in interacting with a computer. 
Touch-screen devices are an exception, though they are not very common in office settings. 

\emph{Uniqueness} of keystroke dynamics is tightly related with its \emph{Circumvention Difficulty}. 
Working systems have been developed that exhibit lower than $0.01\%$ false negative rates \cite{ahmed2014biometric}.

\emph{Permanence} is quite weak since a given user's keystroke dynamics can change significantly  
over the course of a single session, based on specific actions, e.g., typing with both hands or one (pecking).
User fatigue is another factor negatively influencing permanence. 

\emph{Unobtrusiveness} is quite good, {\bf as long as} the user keeps typing naturally.
However, except for a few jobs that require constant typing, most office-setting users
do not engage in continuous typing throughout a typical workday.  In particular, [A2]--[A8]
preclude data collection and would likely result in erroneous de-authentication.

\emph{Collectability} of keystroke dynamics is trivial and excellent, since the biometric itself is defined in the context of the
pre-existing human-computer interaction.

\emph{Cost-Effectiveness} is ideal. Because no hardware instrumentation is needed (only software support), 
systems based on keystroke dynamics are clearly the most {cost-effective} of all methods we consider.

\subsection{\acro~}
We finally assess the proposd \acro\ techniques which attains continuous authentication based on the 
posture-pattern biometric. 

{\em Universality} is very good, as the only user requirement is to be seated, which (perhaps sadly) 
accounts for most of a typical office user's workday and thus applies to the overwhelming majority of users. 
It also correctly functions in cases [A1]--[A9], i.e., users are not unnecessarily de-authenticated.

{\em Uniqueness} is quite high, based on our results (see below) that demonstrate $91\%$ accuracy 
of user identification using measurements collected from individual users over several days.

\emph{Permanence} is, for now, the weak point of \acro. As discussed in Section \ref{sec:results},
based on our initial experiments, it is not satisfactory. However, we believe that further
longer-term studies might change this result.

{\em Unobtrusiveness} is easily achieved due to \acro\ passively collecting data while the user is seated, and
requires no behavior modification or any other burden.

{\em Circumvention Difficulty} is excellent, based on our discussion in Section \ref{sec:threat}.
This is clearly the case for the \emph{casual} adversary, given our $0.33\%$ false positive rate for all subjects. 
We also argue that the burden and cost of creating and deploying a replica in the case of the \emph{determined} 
adversary makes the attack impractical.

{\em Collectability} is very easy: data is obtained using a standard office chair instrumented 
with 16 ultra-thin  force-resistive pressure sensors, which are used to discretize and collect posture pattern data. 

{\em Cost-effectiveness} was one of our key design goals and we believe that it is fairly good. 
All equipment necessary for us to intrument a single office chair prototype costed $\$275$.
However, scale substantially drives down the price: based on cost estimates from hardware providers, 
instrumenting $50$ office chairs would cost $\$150$ per unit, for a total of $\$7,500$. Additionally, the components are easy to acquire. The a401 sensors were purchased from Tekscan\footnote{See: \url{http://tekscan.com} for more information and pricing.}, and the remaning components: the Arduino, breadboard and jumper wires, can be found at any consumer electronics store\footnote{see:\url{http://www.frys.com} for pricing}.
%%GTS: add citations: where would you buy it?
%%          Also, see paragraph at the end of Section \ref{methodology}!
%%         In the rebuttal, you claimed that we can bring it down to $150!!! If so, change here and change the table below too!
%%

	\begin{table*}[!htb]
	\centering 
	\caption{\small Comparison of Continuous Authentication Systems}
	\label{tab:comp}
	%{%\setlength{\extrarowheight}{10pt}
	\begin{tabular}{||c|c|c|c|c|c|c|c||}
	\hline \cline{1-8}
&&&&&&&\\
	\small{\bf System} & \small{\bf Universal} & \small{\bf Unique}& \small{\bf Permanent} & \small{\bf Unobtrusive} & 
	\small{\bf Difficult to Circumvent} & \small{\bf Collectible} & \small{\bf Cost per unit}
	\\ 
&&&&&&&\\ \hline \cline{1-8} 
&&&&&&&\\
%%
%%GTS: I replaced \Checkmark with {\bf WEAK} in the permanence column of \acro. This is correct, right?
%%	 
%%
	\acro	& \Checkmark 	&	\Checkmark & {\bf WEAK} & \Checkmark & \Checkmark & \Checkmark& $\$150$ 
	\\  \hline
&&&&&&&\\
	Gaze Tracking& \Checkmark 	& 	\Checkmark& {\bf X} &\Checkmark& \Checkmark &\Checkmark&$\$2K-\$5K$ 
	\\  \hline
&&&&&&&\\
	Pulse-response	& \Checkmark 	&	\Checkmark& \Checkmark &\Checkmark & \Checkmark & \Checkmark&  Unknown
	\\  \hline
&&&&&&&\\
	ZEBRA	&  \Checkmark	& 	\Checkmark&\Checkmark  & {\bf X} & {\bf X} &\Checkmark&$\$100$ 
	\\  \hline\
&&&&&&&\\
	Keystroke Dynamics	&  \Checkmark	& 	\Checkmark&\Checkmark  & \Checkmark & \Checkmark&\Checkmark& Free
	\\ \hline \cline{1-8}
%&&&&&&&\\
	\end{tabular}
	\end{table*}

\section{Methodology}
\label{sec:method}
This section describes our initial \acro\ prototype design, experimental setup, procedures, subject parameters as well as
classifiers used for data analysis.

\subsection{Prototype Design}
\begin{figure*}[t!]
\fbox{\centering
\includegraphics[height=3.6in,width=2.1\columnwidth]{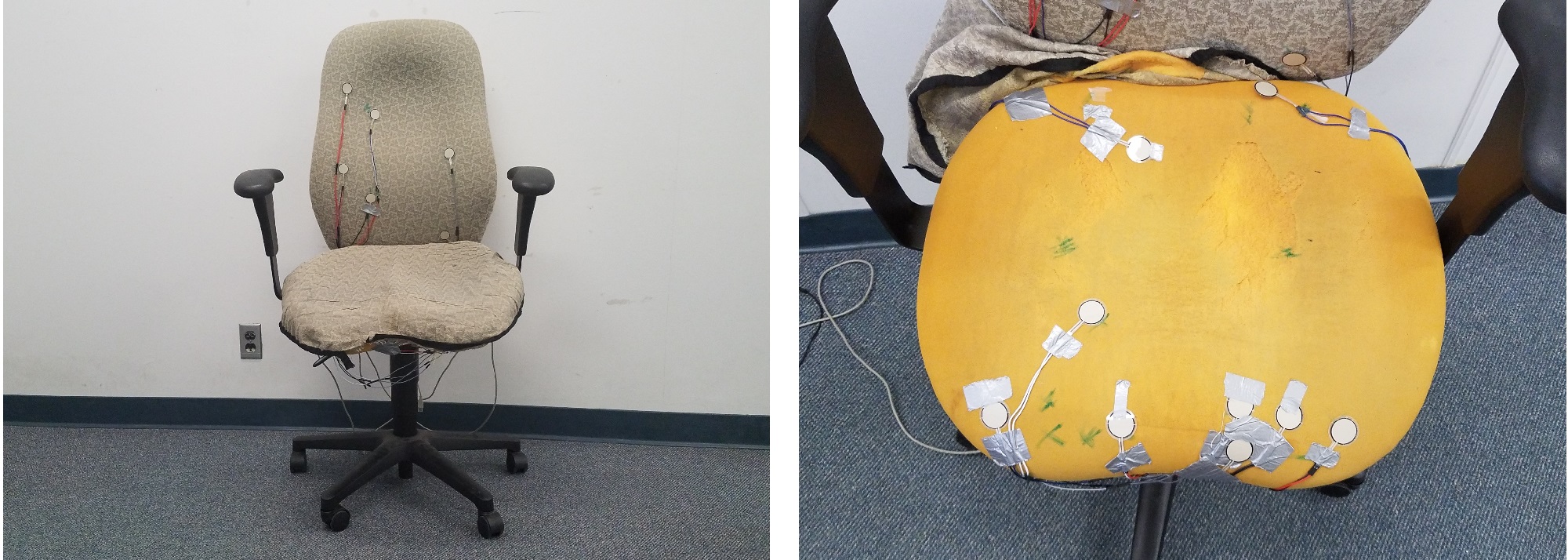}}
\caption{{\small The \acro~ prototype chair: (a) as seen by user, and (b) uncovered seat-bottom sensor placements.} }
\label{fig:chair}
\end{figure*} 
\begin{figure}[h]
\fbox{\centering
\includegraphics[height=3.6in,width=\columnwidth]{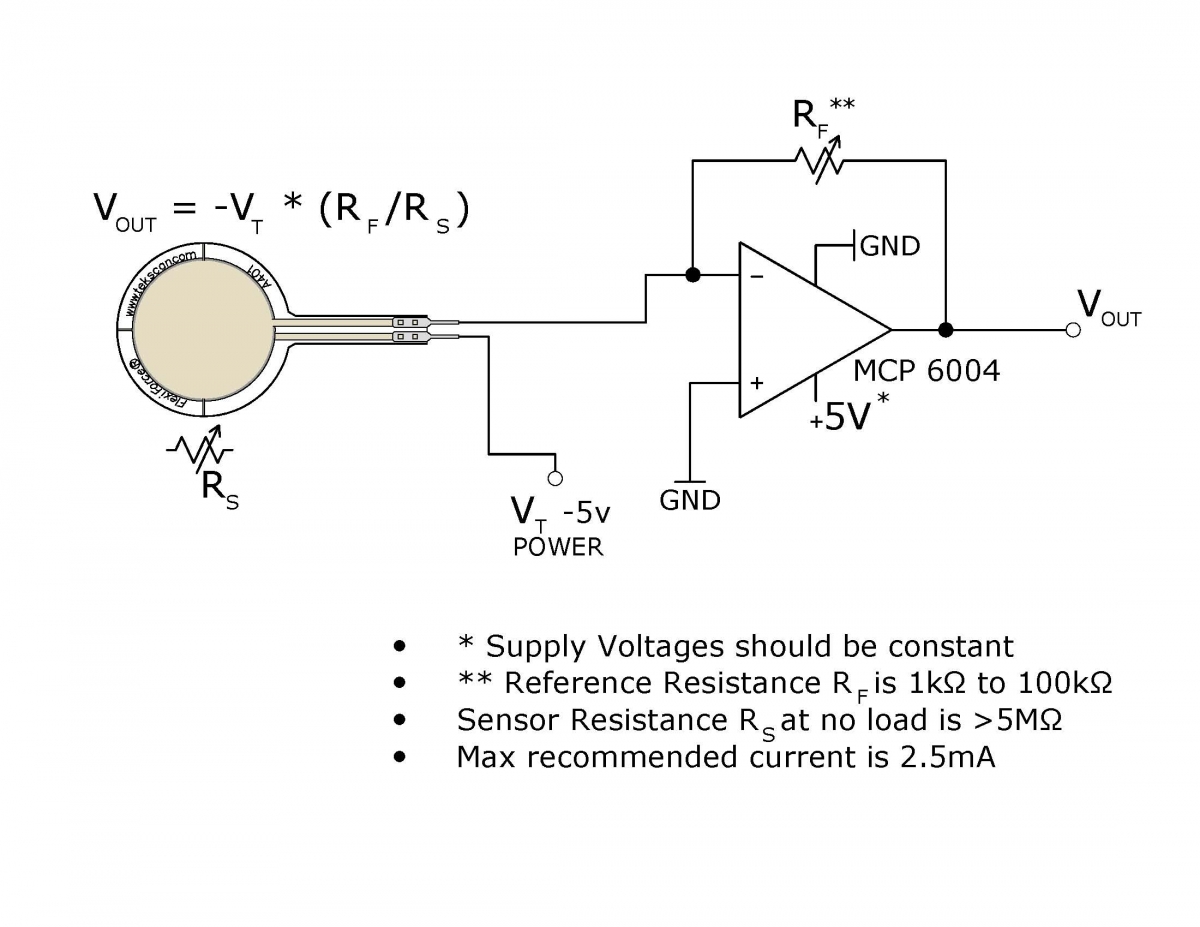}}
\caption{{\small } Recommended circuit for the Tekscan Flexiforce A401 sensor.}
\label{fig:circ}
\end{figure} 
To demonstrate  viability and facilitate ease of experimentation, we built the \acro~ prototype by 
modifying a standard inexpensive office chair with commodity (off-the-shelf) sensor components. 
Figure \ref{fig:chair}(a) shows the prototype chair, and Figure \ref{fig:chair}(b) focuses on 
the placement of sensors across the seat and back of the chair. Our sensor placement was 
guided by the experience of Mutlu et al. \cite{mutlu2007robust}.

\begin{figure}[h]
\fbox{\centering
\includegraphics[height=3.6in,width=1\columnwidth]{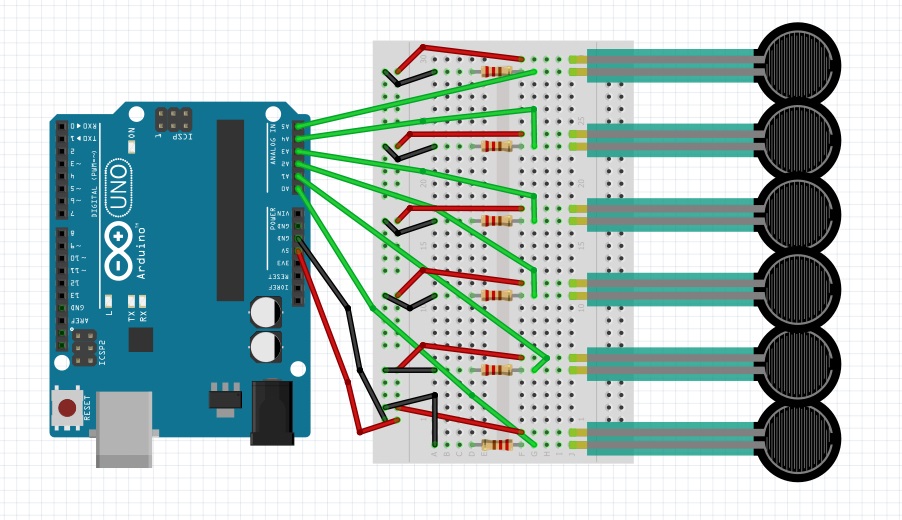}}
\caption{{\small Fritzing diagram for the 6-sensor Arduino.} }
\label{fig:schem}
\end{figure} 

The prototype consists of three components:
\begin{compactenum}
\item One 2003/2004 Hon Mid-Back Task Chair.\footnote{See \url{https://www.hon.com}.}
\item Sixteen (16) Tekscan Flexiforce A401 Large Force Sensing Resistors.$^6$
The recommended circuit for pressure data acquisition is illustrated in Figure \ref{fig:circ}.
\item Two Arduino 101 modules\footnote{See \url{https://www.arduino.cc/en/Main/ArduinoBoard101}.}, 
one of which is connected to $6$ A401 resistors, as shown in Figure \ref{fig:schem}. 
The other module is connected to the remaining 10 sensors in a similar configuration, augmented with an 
analog multiplexer in its $6$-th analog port, in order to support the use of $10$ sensor inputs.
\end{compactenum}
Acquired measurements are sent from the Arduino to a commodity desktop PC for collection and evaluation. 
Arduinos are connected to the desktop via USB cables. Obviously, in a real office setting, having wires running 
between the chair and the computer would be highly undesirable. We expect that either Bluetooth or WiFi would
be used instead.  

As mentioned earlier in the paper, our total instrumentation cost of $\$275$ was for the initial single-chair 
prototype, created to demonstrate feasibility and for testing via user experiments. For a medium-size office with 
50 chairs, the price can be cut significantly, to approximately $\$150$ per instrumented chair due to the reduced cost of bulk ordering the Tekscan A401$^6$. 
%%GTS: Add more info! Where would you buy it, what would be the price! 

\subsection{Data Collection Procedure}
To collect data in a realistic setting, rather than bringing subjects to an unfamiliar office
and encountering complications cited by Yamada, et. al \cite{yamada2009soft} in collecting posture data in a lab setting, 
we brought the prototype instrumented chair to the subjects' workplace.  Each subject was briefed on the nature of the 
experiment, and was asked to sit naturally. Subjects allowed us to swap out their office chair with the prototype, and 
continued their normal work activities while sitting on the latter. We collected posture data in rounds of $10$ minutes per 
subject. $17$ subjects participated in two collection sessions over the course of several days. and $13$ 
subjects participated in a single session only. We sampled subjects in order to accommodate typical day-to-day 
fluctuations in mood and posture, e.g., one session in the morning, and the other -- shortly after lunch, on a different day.

A total of $30$ subjects were recruited primarily from the graduate student population of a large public university. 
Because of this, overwhelming majority ($27$ out of $30$) were between the ages of $22$ and $30$, while the
the remaining $3$ were somewhat older faculty and staff. The subjects' gender break-down was: $10$ female and $20$ male.

Finally, despite its somewhat ungainly appearance (as shown in Figure \ref{fig:chair}), the prototype chair
is rather comfortable for sitting and none of the subjects expressed any unease or discomfort during
the data collection phase.

\subsection{Features}
\begin{figure}[h]
\fbox{\centering
\includegraphics[height=3.0in,width=1\columnwidth]{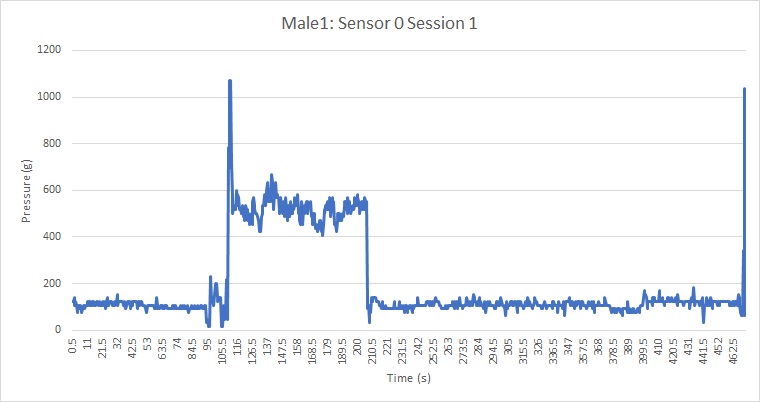}}
\caption{{\small Raw sensor data for demonstrating a posture shift.} }
\label{fig:shift}
\end{figure}
We collected data in the form of $1,200$ sample time-series reflecting the force exerted on each of the $16$ 
pressure sensors captured each $0.5$ seconds over a $10$-minute session, for a total of $19,200$ samples 
per subject, per session. Figure \ref{fig:shift} shows an example of pressure  data collected from a single sensor 
over the course of a single session. Note that the subject's posture changes part-way through the session. 
The subject then remains in the new posture for several minutes, and later shifts again to a posture similar to the first. 

For continuous authentication, we treat the first $5$ minutes of each session as a training phase, and evaluate the 
subject on the next $10$ seconds. If that data is accepted as valid, it is included in the training set for the next 
$10$ second evaluation window.
 
\subsection{Classification Algorithm}
Since we are dealing with a fairly commonplace time series clustering problem, there are many 
well-known candidate techniques. We compared three most popular classification
algorithms to determine the one that provides the best results.

\noindent\emph{Random Forest (RF):} 
we found that it consistently yields the best results. It produces 
precise, accurate results, closely clustered for all subjects. Both false negative and false positive rates 
are acceptably low in cross-validation of user data, as discussed in more detail below.

\noindent\emph{K-Nearest Neighbors (KNN):} 
we tested the KNN classifier for $k=1,\;3$ and $5$ using euclidean 
distance. KNN is a simple lazy classifier that is quite effective in many settings. However, for our 
specific classification needs, it did not perform as well as RF.

\noindent\emph{Support Vector Machine (SVM):}
For each subject, we trained a single binary classifier in a one-against-one case. The final prediction is 
determined by voting.  While SVM provided extremely consistent and highly accurate results for some users, 
it did not perform as well RF, on average. It also had a few outliers with unacceptably high false positive rates.

\section{Results}
\label{sec:results}
We present results for two classifiers: one for identification and the other -- for authentication. The former is based on  
RF and provides verification of a one-to-$n$ match of a sample of a known user against every sample in a database. 
The authentication classifier is also RF-based and provides verification of a one-to-one match of a 
sample of unknown origin against that of a single known user.

We divide the results into two parts: (1) those from a single test set, which show distinguishing 
capabilities of \acro, and (2) those from applying the classifier to data sampled over time across multiple user 
sessions. The latter are used to test \emph{permanence} of our \acro~ prototype.

\begin{figure*}[h!]
\fbox{\centering\includegraphics[height=2.6in,width=\columnwidth * 2]{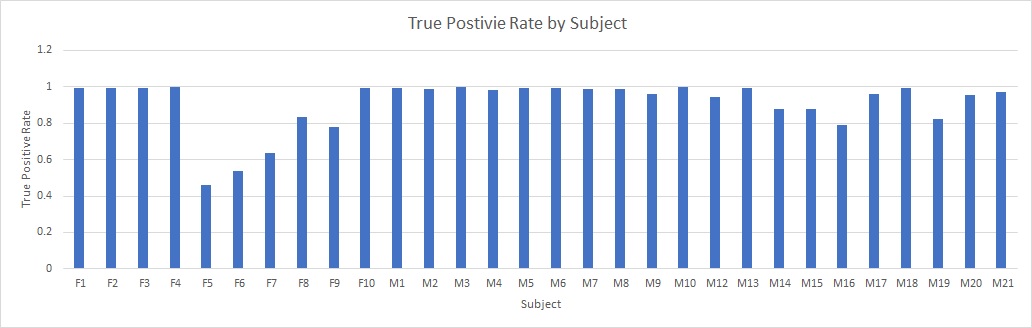}}
\caption{{\small  Identification classifier results: True Positive Rate for each subject is obtained by 
performing 10 times stratified 10-fold cross-validation.}}
\label{fig:TRP}
\end{figure*}

\begin{figure*}[h!]
\fbox{\centering\includegraphics[height=2.6in,width=\columnwidth *2]{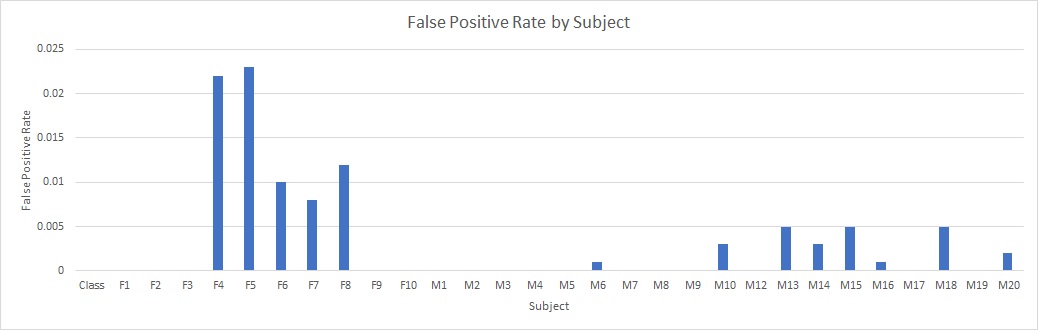}}
\caption{{\small  False Positive Rate for Random Forest cross-validation on all subject data.}}
\label{fig:FPR}
\end{figure*}

\begin{figure*}[h!]
\fbox{\centering\includegraphics[height=2.6in,width=\columnwidth *2]{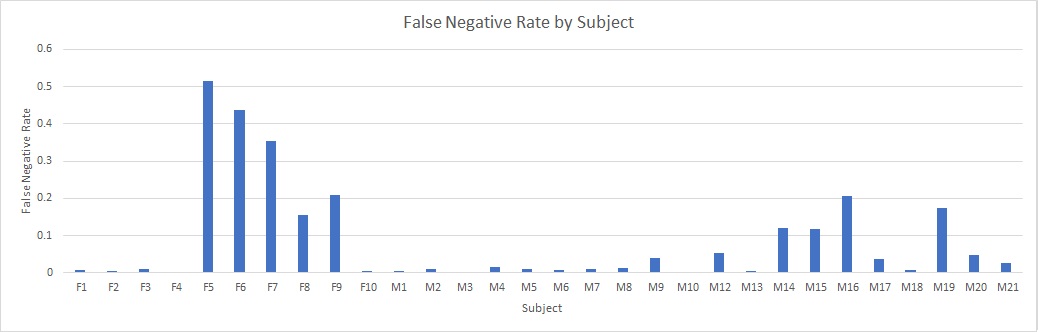}}
\caption{{\small False Negative rate for Random Forest cross-validation on all subject data.}}
\label{fig:FNR}
\end{figure*}

Identification is a classification problem across many classes. Our RF classifier is ideal for this -- it 
achieves, on average, $91.0\%$  true positive rate, as shown in Figure \ref{fig:TRP}. We also achieve average 
false positive rate of $0.33\%$, as shown in Figure \ref{fig:FPR}, and $8.68\%$ average false 
negative rate, shown in Figure \ref{fig:FNR}. 

As results indicate, \emph{permanence} is problematic in our prototype. When evaluating user data 
across multiple sessions, the RF classifier yielded only a $22\%$ true positive rate, a false positive 
rate of $5.2\%$, and a $72.7\%$ false negative rate. However, we do not view this as a major 
weakness of either \acro or its initial prototype. 
Instead, we consider this to be a  motivation for a more in-depth longitudinal study. 

\section{Discussion}
\label{sec:discuss}

We now assess \acro\ in the context of design goals for an ideal de-authentication system 
outlined in Section \ref{sec:bg}.

\acro~ was designed with the emphasis on \emph{minimal extra components and monetary cost}. 
The use of an instrumented chair does require specialized hardware. However, it does not impose any 
behavioral requirements on the user. Furthermore, the per-unit cost of $\$150$ (at scale of about $50$) 
is reasonable in the context of other posture-based techniques, which can easily cost thousands of 
dollars, e.g., \cite{mutlu2007robust}. 

We claim that \emph{maximal user transparency} is achieved by \acro~ because of the ubiquitous nature of sitting 
in office or workplace settings. In fact, over $70\%$ of the workforce in a traditional office setting spend upwards of 
$6$ hours a day seated \cite{ryan2011sitting}. Enrollment, authentication and de-authentication phases of \acro all occur 
transparently while the user is seated and engaged in normal workflow activities. Because of this, there are no 
behavioral modifications required from the user to participate in \acro and no need for modifying everyday activities.

\emph{Quick detection of activities requiring de-authentication} is trivial in \acro. A user who engages in any activity 
covered by [A9] is  de-authenticated as soon as a single collection window passes. Though in the initial prototype 
implementation this window was set to $10$ seconds, it can be adjusted up or down.

As evidenced by the average $91.0\%$ rate of accuracy of user identification, false positives would only occur in 
exceptional circumstances, which satisfies the \emph{minimal insult rate} design goal. This holds during most
typical office activities [A1-A7] that are typically performed while the users is seated. However, if a user leaves the
chair to grab something nearby [A8] and spends over 10 seconds away, potentially erroneous de-authentication can
occur. 

\acro prototype achieves a very low average false negative rate of $0.33\%$. Even if a \emph{casual} 
adversary succeeds once, the probability of two successive successful impersonations is only $0.11\%$. 
We believe that this satisfies the goal of having a \emph{minimal fraud rate}. 

\subsection{Ethical Concerns}
Unfortunately, the use of \acro~ for de-authentication system also triggers potentially negative implications 
for user privacy. Unlike traditional single-session authentication techniques, \acro~ is inherently tied to the user's 
physical presence. This has the unintended consequence of leaking whether or not the user is physically at the 
office or desk. This information could be used by the unscrupulous management to micro-manage and/or 
audit the time an employee spends at the workstation, the frequency with which they get up, how long they are 
gone, and other very personal details. 

This type of privacy leakage is not unique to \acro. Indeed, it is common to most continuous authentication and de-authentication
methods. In general, details corresponding to each time when the user quits acting in the expected manner are leaked. 
(For example, with keystroke dynamics, information is leaked whenever the user stops typing.) We certainly do not 
recommend or endorse the use of \acro in such ethically dubious ways.

All experiments described in this paper were duly authorized by the 
Institutional Review Board (IRB) of the authors' employer, well ahead of the actual commencement of the study. 
The level of review was: Exempt, Category II. No sensitive data was collected during the experiments and minimal 
identifying information was retained. In particular, no subject names, phone numbers or other personally identifying 
information (PII) was collected. All obtained data was stored pseudonymously.

%
%before conlusion there's future work: need to create a small section on future work
%future work: longitudinal study of permanence (understand permanence better) 
%although our results ar enot *great* the expeirments were narrow and aimed at uniqueness and aimed at deauthentication 
%plan to do a natural over the course of the day study e.g. grab a subject over the course of a typical workday (field trial) to estimate practical FP rate
%Also gonna try to study the attack suceptability; try people to emulate 
%how obstrusive, cheap and effective a butt simaculrum 
\section{Future Work}
\label{sec:FW}
There are several directions for future work.

First, as mentioned above, our initial permanence results are weak. We plan to conduct a larger-scale, longer-term
(longitudinal) study, obtaining multiple measurement sessions from each subject. This would yield more concrete 
permanence results and lead to a better understanding of the posture pattern biometric as  a whole.

Second, we intend to evaluate accuracy of \acro in its typical use-case. For this field study, we intend to have the 
subjects replace their office chair with our prototype for an entire workday. We would use this data to obtain the
rate of both false negatives and positivies, throughout the day, as well as measure associated user burden.

Next, we plan to evaluate attack vectors outlined in section \ref{sec:threat}, starting with a \emph{casual} adversary. 
This will entail recruiting pairs of subjects with similar physical characteristics, and training them to impersonate each other's 
posture patterns.  % This will establish a more realistic understanding of the rate of false positives for \acro. 
Finally, we explore the attacks by a \emph{determined} adversary. For this we need to construct a contraption that imitates 
the victim's posture pattern. 
% We will evaluate feasible such an attack would be in the workplace, as well as how cost-effective and non-obtrusive the prosthetic is.

\section{Conclusion}
\label{sec:conc}
In summary, this paper proposed and described a new biometric -- \acro -- based on seated posture patterns. We built
and experimented with a prototype implementation of \acro.
Furthermore, experimental results show that posture pattern biometric captures a unique combination of physiological and
behavioral traits. We found that users can be identified with, on average, $91\%$ accuracy from a population of $30$. 
We also believe that it is infeasible for a \emph{casual} adversary to circumvent \acro by impersonation of the victim's posture 
patterns. We also argue that physical  and logistical burdens of fabricating and deploying an accurate mold (replica) of 
the victim's relevant body parts make circumvention very challenging even for the \emph{determined} adversary.
Finally, we provided a thorough comparison of several prominent modern biometric-based techniques for continuous 
authentication. 

\bibliographystyle{IEEEtranS}
\bibliography{assentication}
\end{document}